\documentclass[12pt,letterpaper]{article}
\usepackage{amsmath,amssymb,array,calc,rotating,epsfig,psfrag, amscd}
\usepackage{color}
\usepackage[
      colorlinks=true,
      urlcolor=blue,    
      filecolor=blue,     
      citecolor=red,
      pdfstartview=FitV,
       bookmarksopen=true    
      ]{hyperref}
\usepackage[left=2cm,top=2cm,right=2cm,nohead]{geometry}

\newcommand{\nc}{\newcommand}

\nc{\ra}{\rightarrow} 
\nc{\lra}{\leftrightarrow} 
\nc{\Ra}{\Rightarrow} 
\nc{\LRa}{\Leftightarrow} 
\nc{\blp}{{\big (}}
\nc{\brp}{{\big )}}
\nc{\Blp}{{\Big (}}
\nc{\Brp}{{\Big )}}
\nc{\bglp}{{\bigg (}}
\nc{\bgrp}{{\bigg )}}
\nc{\Bglp}{{\Bigg (}}
\nc{\Bgrp}{{\Bigg )}}
\nc{\slb}{{\rm [}}
\nc{\srb}{{\rm ]}}
\nc{\bslb}{{\rm \big [}}
\nc{\bsrb}{{\rm \big ]}}
\nc{\Bslb}{{\rm \Big [}}
\nc{\Bsrb}{{\rm \Big ]}}
\def\al{\alpha}

\def\eps{\epsilon}
\nc{\veps}{\varepsilon}
\def\gam{\gamma}

\def\lam{\lambda}
\def\om{\omega}
\nc{\vphi}{\varphi}
\def\tha{\theta}
\def\sig{\sigma}

\def\Gam{\Gamma}

\def\Om{\Omega}
\def\Sig{\Sigma}

\nc{\myvspace}{\rule[-1em]{0pt}{2.5em}}
\nc{\bea}{\begin{eqnarray}}
\nc{\eea}{\end{eqnarray}}
\nc{\be}{\begin{equation}}
\nc{\ee}{\end{equation}}
\nc{\barr}{\begin{array}}
\nc{\earr}{\end{array}}
\nc{\cA}{{\cal A}}
\nc{\cB}{ \cal B}

\def\cD{{\cal D}}

\nc{\cF}{{\cal F}}
\nc{\cG}{{\cal G}}
\def\cH{{\cal H}}
\def\cI{{\cal I}}
\def\cJ{{\cal J}}
\def\cK{{\cal K}}
\nc{\cL}{{\cal L}}
\nc{\cM}{{\cal M}}
\def\N{{\cal N}}
\def\cN{{\cal N}}

\nc{\cQ}{{\cal Q}}
\nc{\cR}{{\cal R}}

\def\cV{{\cal V}}
\def\cV{{\cal V}}

\def\cZ{{\cal Z}}
\nc{\cQd}{\cQ^{\dagger}}
\nc{\cRd}{\cR^{\dagger}}
\nc{\BB}{{\mathbb B}}
\nc{\CC}{{\mathbb C}}
\nc{\DD}{{\mathbb D}}
\nc{\EE}{{\mathbb E}}
\nc{\FF}{{\mathbb F}}
\nc{\GG}{{\mathbb G}}
\nc{\HH}{{\mathbb H}}
\nc{\JJ}{{\mathbb J}}
\nc{\RR}{{\mathbb R}}
\nc{\PP}{{\mathbb P}}
\nc{\QQ}{{\mathbb Q}}
\nc{\ZZ}{{\mathbb Z}}
\nc{\calone}{{\mathbb 1}}

\nc{\half}{\frac{1}{2}}
\nc{\qrt}{\frac{1}{4}}
\nc{\del}{\partial}

\nc{\delbar}{\bar\partial}
\nc{\thalf}{\frac{t}{2}}
\nc{\Spin}{\operatorname{Spin}}
\nc{\SO}{\operatorname{SO}}

\nc{\Sp}{{\rm Sp}}
\nc{\com}[2]{{ \left[ #1, #2 \right] }}
\nc{\acom}[2]{{ \left\{ #1, #2 \right\} }}
\nc{\rr}{\rightarrow}
\nc{\p}{\partial}
\nc{\LT}{{\LL_\T}}
\nc{\Tr}{{\rm Tr}}
\nc{\tr}{{\rm tr}}
\def\com#1#2{{ \left[ #1, #2 \right] }}
\def\acom#1#2{{ \left\{ #1, #2 \right\} }}
\nc{\Adag}{A^{\dagger}}
\nc{\AdagI}{A^{\dagger I}}
\nc{\AdagJ}{A^{\dagger J}}
\nc{\AdagK}{A^{\dagger K}}
\nc{\AdagL}{A^{\dagger L}}
\nc{\AdagM}{A^{\dagger M}}
\nc{\Bdag}{B^{\dagger}}
\nc{\BdagI}{B^{\dagger}_I}
\nc{\BdagJ}{B^{\dagger}_J}
\nc{\BdagK}{B^{\dagger}_K}
\nc{\BdagL}{B^{\dagger}_L}
\nc{\BdagM}{B^{\dagger}_M}
\nc{\Cdag}{C^{\dagger}}
\nc{\CdagI}{C^{\dagger I}}
\nc{\CdagJ}{C^{\dagger J}}
\nc{\CdagK}{C^{\dagger K}}
\nc{\Ddag}{D^{\dagger}}
\nc{\DdagI}{D^{\dagger I}}
\nc{\DdagJ}{D^{\dagger J}}
\nc{\DdagK}{D^{\dagger K}}
\nc{\ttha}{\tilde{\theta}}
\nc{\tphi}{\tilde{\phi}}
\nc{\tsig}{\tilde{\sig}}
\nc{\tom}{\tilde{\om}}
\nc{\tlam}{\tilde{\lam}}
\nc{\tSig}{\widetilde{\Sig}}
\nc{\tPhi}{\tilde{\Phi}}
\nc{\tPhibar}{\ol{\tPhi}}
\nc{\tPi}{\tilde{\Pi}}
\nc{\tpsi}{\tilde{\psi}}
\nc{\tPsi}{\tilde{\Psi}}
\nc{\tgam}{\tilde{\gam}}
\nc{\tGam}{\tilde{\Gam}}
\nc{\tzeta}{\tilde{\zeta}}
\nc{\tZeta}{\tilde{\Zeta}}
\nc{\teta}{\tilde{\eta}}
\nc{\teps}{\tilde{\eps}}
\nc{\tEta}{\tilde{\Eta}}
\nc{\tchi}{\tilde{\chi}}
\nc{\tChi}{\tilde{\Chi}}
\nc{\Xit}{\tilde{\Xi}}
\nc{\tb}{\tilde b}
\nc{\tc}{\tilde c}
\nc{\te}{\tilde e}
\nc{\tf}{\tilde f}
\nc{\tg}{\tilde g}
\nc{\tj}{\tilde j}
\nc{\tp}{\widetilde{p}}
\nc{\tq}{\widetilde{q}}
\nc{\ts}{{\tilde s}}
\nc{\tu}{{\tilde u}}
\nc{\tv}{{\tilde v}}
\nc{\tw}{{\tilde w}}
\nc{\tx}{{\tilde x}}
\nc{\ty}{{\tilde y}}
\nc{\tz}{\tilde z}
\nc{\tA}{{\tilde A}}
\nc{\tAbar}{{\ol \tA}}
\nc{\tB}{{\widetilde B}}
\nc{\tC}{{\widetilde C}}
\nc{\tD}{{\widetilde D}}
\nc{\tE}{{\widetilde E}}
\nc{\tG}{{\widetilde G}}
\nc{\tH}{{\widetilde H}}
\nc{\tJ}{{\widetilde J}}
\nc{\tJbar}{{\ol {\tilde J}}}
\nc{\tK}{{\widetilde K}}
\nc{\tL}{{\widetilde L}}
\nc{\tM}{{\widetilde M}}
\nc{\tN}{{\widetilde N}}
\nc{\tP}{{\widetilde P}}
\nc{\tQ}{{\widetilde Q}}
\nc{\tR}{{\widetilde R}}
\nc{\tS}{\widetilde{S}}
\nc{\tF}{\tilde{{\cal F}}}
\nc{\tX}{\widetilde{X}}
\nc{\tcZ}{\tilde{\cZ}}
\nc{\tcZbar}{\ol{\tcZ}}
\nc{\hb}{\hat b}
\nc{\hc}{\hat c}
\nc{\hd}{\hat d}
\nc{\he}{\hat e}
\nc{\hf}{\hat f}
\nc{\hg}{\hat g}
\nc{\hh}{\hat h}
\nc{\hp}{\hat p}
\nc{\hs}{\hat s}
\nc{\hv}{\hat v}
\nc{\hw}{\hat w}
\nc{\hx}{\hat x}
\nc{\hy}{\hat y}
\nc{\hz}{\hat z}
\nc{\zhat}{\hat z}
\nc{\hA}{\widehat{A}}
\nc{\hE}{\widehat{E}}
\nc{\hF}{\widehat{F}}
\nc{\hH}{\widehat{H}}
\nc{\hJ}{\widehat{J}}
\nc{\hK}{\widehat{K}}
\nc{\hL}{\widehat{L}}
\nc{\hM}{\widehat M}
\nc{\hN}{\widehat{N}}
\nc{\hV}{\widehat V}
\nc{\hcV}{\widehat \cV}

\nc{\ha}{\widehat \alpha}
\nc{\hphi}{\hat{\phi}}
\nc{\hpsi}{\hat{\psi}}
\nc{\hgam}{\hat{\gam}}
\nc{\hPhi}{\hat{\Phi}}
\nc{\hPsi}{\hat{\Psi}}
\nc{\hGam}{\hat{\Gam}}
\nc{\omhat}{\hat{\om}}

\nc{\w}{\wedge}

\nc{\vb}{\vec b}
\nc{\vc}{\vec c}
\nc{\vd}{\vec d}
\nc{\ve}{\vec e}
\nc{\vf}{\vec f}
\nc{\vg}{\vec g}
\nc{\vh}{\vec h}
\nc{\vp}{\vec p}
\nc{\vq}{\vec q}
\nc{\vr}{\vec r}
\nc{\vs}{\vec s}
\nc{\vv}{\vec v}
\nc{\vw}{\vec w}
\nc{\vx}{\vec x}
\nc{\vy}{\vec y}
\nc{\vz}{\vec z}

\nc{\ol}{\overline}
\nc{\abar}{\ol{a}}
\nc{\bbar}{\ol{b}}
\nc{\cbar}{\ol{c}}
\nc{\dbar}{\ol{d}}
\nc{\ebar}{\ol{e}}
\nc{\ibar}{\ol{\imath}}
\nc{\jbar}{\ol{\jmath}}
\nc{\kbar}{\ol{k}}
\nc{\lbar}{\ol{l}}
\nc{\mbar}{\ol{m}}
\nc{\nbar}{\ol{n}}
\nc{\pbar}{\ol{p}}
\nc{\qbar}{\ol{q}}
\nc{\ubar}{\ol{u}}
\nc{\vbar}{\ol{v}}
\nc{\wbar}{\ol{w}}
\nc{\xbar}{\ol{x}}
\nc{\ybar}{\ol{y}}
\nc{\zbar}{\ol{z}}

\nc{\Abar}{\ol{A}}
\nc{\Bbar}{\ol{B}}
\nc{\Cbar}{\ol{C}}
\nc{\Dbar}{\ol{D}}
\nc{\Ebar}{\ol{E}}
\nc{\Fbar}{\ol{F}}
\nc{\Jbar}{\ol{J}}
\nc{\Kbar}{\ol{K}}
\nc{\Lbar}{\ol{L}}
\nc{\Mbar}{\ol{M}}
\nc{\Nbar}{\ol{N}}
\nc{\Pbar}{\ol{P}}
\nc{\Qbar}{\ol{Q}}
\nc{\Rbar}{\ol{R}}
\nc{\Sbar}{\ol{S}}
\nc{\Tbar}{\ol{T}}
\nc{\Ubar}{\ol{U}}
\nc{\Vbar}{\ol{V}}
\nc{\Wbar}{\ol{W}}
\nc{\Xbar}{{\overline X}}
\nc{\Ybar}{{\overline Y}}
\nc{\Zbar}{{\overline Z}}
\nc{\cZbar}{{\overline \cZ}}

\nc{\epsbar}{\ol{\epsilon}}
\nc{\lambar}{\ol{\lambda}}
\nc{\zetabar}{\ol{\zeta}}
\nc{\Zetabar}{\ol{\Zeta}}
\nc{\psibar}{\ol{\psi}}
\nc{\Psibar}{\ol{\Psi}}
\nc{\phibar}{\ol{\phi}}
\nc{\Phibar}{\ol{\Phi}}
\nc{\chibar}{\ol{\chi}}
\nc{\mubar}{\ol{\mu}}
\nc{\nubar}{\ol{\nu}}
\nc{\rhobar}{\ol{\rho}}
\nc{\ombar}{\ol{\om}}
\nc{\Ombar}{\ol{\Om}}
\nc{\Deltabar}{\ol{\Delta}}
\nc{\Thetabar}{\ol{\Theta}}
\nc{\xibar}{\ol{\xi}}
\nc{\Xibar}{\ol{\Xi}}

\nc{\Dthbar}{\ol{\rm D3}}

\nc{\gdot}{\dot{g}}
\nc{\xdot}{\dot{x}}
\nc{\ydot}{\dot{y}}
\nc{\phidot}{\dot{\phi}}
\nc{\sinp}{s_{\phi}}
\nc{\cosp}{c_{\phi}}
\nc{\tanp}{t_{\phi}}
\nc{\spone}{s_{\phi_1}}
\nc{\cpone}{c_{\phi_1}}
\nc{\tpone}{t_{\phi_1}}
\nc{\sptwo}{s_{\phi_2}}
\nc{\cptwo}{c_{\phi_2}}
\nc{\tptwo}{t_{\phi_2}}
\nc{\spth}{s_{\phi_3}}
\nc{\cpth}{c_{\phi_3}}
\nc{\tpth}{t_{\phi_3}}
\nc{\calp}{c_{\al}}
\nc{\salp}{s_{\al}}

\nc{\csch}{{\rm csch}}
\nc{\sech}{{\rm sech}}

\nc{\bah}{{\mathbf {\hat{A}}}}
\nc{\bX}{{\mathbf X}}
\nc{\ba}{{\bf a}}
\nc{\bb}{{\bf b}}
\nc{\bc}{{\bf c}}
\nc{\bd}{{\bf d}}
\nc{\bg}{{\bf g}}
\nc{\bk}{{\bf k}}
\nc{\bl}{{\bf l}}
\nc{\bm}{{\bf m}}
\nc{\bn}{{\bf n}}
\nc{\bo}{{\bf o}}
\nc{\bp}{{\bf p}}
\nc{\bq}{{\bf q}}
\nc{\br}{{\bf r}}
\nc{\bs}{{\bf s}}
\nc{\bt}{{\bf t}}
\nc{\bu}{{\bf u}}
\nc{\bv}{{\bf v}}
\nc{\bw}{{\bf w}}
\nc{\bx}{{\bf x}}
\nc{\by}{{\bf y}}
\nc{\bz}{{\bf z}}
\nc{\bom}{{\bf \om}}
\nc{\bombar}{{\mathbf \ombar}}
\nc{\bPhi}{{\bf \Phi}}

\nc{\rma}{{\rm a}}
\nc{\rmb}{{\rm b}}
\nc{\rmc}{{\rm c}}
\nc{\rmd}{{\rm d}}
\nc{\rmg}{{\rm g}}
\nc{\rk}{{\rm k}}
\nc{\rml}{{\rm l}}
\nc{\rmm}{{\rm m}}
\nc{\rmn}{{\rm n}}
\nc{\rmo}{{\rm o}}
\nc{\rmp}{{\rm p}}
\nc{\rmq}{{\rm q}}
\nc{\rmr}{{\rm r}}
\nc{\rms}{{\rm s}}
\nc{\rmt}{{\rm t}}
\nc{\rmu}{{\rm u}}
\nc{\rmv}{{\rm v}}
\nc{\rmw}{{\rm w}}
\nc{\rmx}{{\rm x}}
\nc{\rmy}{{\rm y}}
\nc{\rmz}{{\rm z}}
\nc{\Ffour}{{F^{(4)}}}
\nc{\Ffive}{{F^{(5)}}}

\nc{\dal}{\dot{\al}}
\nc{\thadot}{\dot{\tha}}
\nc{\thab}{\bar{\theta}}
\nc{\thal}{\theta^{\al}}
\nc{\thdal}{\bar{\theta}^{\dal}}

\nc{\thsigthm}{\tha \sigma^m \thab}
\nc{\thsigthn}{\tha \sigma^n \thab}

\nc{\Dal}{D_{\al}}
\nc{\Ddal}{\bar{D}_{\dal}}
\nc{\CDal}{{\cal D}_{\al}}
\nc{\CDdal}{\bar{\cal D}_{\dal}}

\nc{\eq}[1]{(\ref{#1})}
\nc{\non}{\nonumber}

\nc{\equ}{{\rm eq}}
\def\Im{{\rm Im ~}}

\def\Omp{{\rm \Om'}}

\def\Re{{\rm Re ~}}

\nc{\AdS}{{\rm AdS}}
\nc{\vol}{{\rm vol}}
\nc{\Ainf}{A_{\infty}}
\nc{\End}{{\rm End}}
\nc{\Ext}{{\rm Ext}}
\nc{\IIB}{{\rm IIB}}
\nc{\Ad}{{\rm Ad}}
\nc{\IIA}{{\rm IIA}}
\nc{\Dslash}{\ensuremath \raisebox{0.025cm}{\slash}\hspace{-0.32cm} D}
\nc{\cDslash}{\ensuremath \raisebox{0.025cm}{\slash}\hspace{-0.32cm} \cD}
\nc{\no}{\!:\!\!}
\nc{\ointdz}{\oint\frac{dz}{2\pi i}}
\nc{\ointdzone}{\oint\frac{dz_1}{2\pi i}}
\nc{\ointdztwo}{\oint\frac{dz_2}{2\pi i}}
\nc{\ointdzb}{\oint\frac{d\zbar}{2\pi i}}
\nc{\ointdzbone}{\oint\frac{d\zbar_1}{2\pi i}}
\nc{\ointdzbtwo}{\oint\frac{d\zbar_2}{2\pi i}}
\nc{\dz}{\frac{dz}{2\pi i}}
\nc{\dzb}{\frac{d\zbar}{2\pi i}}
\nc{\bpm}{\begin{pmatrix}}
\nc{\epm}{\end{pmatrix}}
 \nc{\bitem}{\begin{itemize}}
 \nc{\eitem}{\end{itemize}}

\definecolor{cardinal}{rgb}{0.6,0,0}
\definecolor{darkgreen}{rgb}{0,0.5,0}
\definecolor{golden}{rgb}{0.92, 0.7, 0}
\definecolor{midnight}{rgb}{0, 0, 0.5}
\definecolor{darkblue}{rgb}{0.2, 0, 0.8}

\begin{document}
\begin{center}
\vskip 2 cm

{\Large \bf Supersymmetric Consistent \\ Truncations of IIB  on $T^{1,1}$}
\vskip 1.25 cm 
{Iosif Bena, Gregory Giecold, Mariana Gra\~na, \\
Nick Halmagyi and Francesco Orsi}\\
\vskip 5mm
Institut de Physique Th\' eorique, \\
CEA/Saclay, CNRS-URA 2306,  \\
Orme des Merisiers, F-91191 Gif sur Yvette, France
\vskip 1 cm
iosif.bena, gregory.giecold, mariana.grana, \\
nicholas.halmagyi, francesco.orsi@cea.fr
\vskip 1 cm
\end{center}

\begin{abstract}

We study consistent Kaluza-Klein reductions of type IIB supergravity on $T^{1,1}$ down to five-dimensions. We find that the most general reduction containing singlets under the global $SU(2)\times SU(2)$ symmetry of $T^{1,1}$ is $\N=4$ gauged supergravity coupled to three vector multiplets with a particular gauging due to topological and geometric flux. Key to this reduction is several modes which have not been considered before in the literature and our construction allows us to easily show that the Papadopoulos - Tseytlin ansatz for IIB solutions on $T^{1,1}$ is a consistent truncation. This explicit reduction provides an organizing principle for the linearized spectrum around the warped deformed conifold as well as the baryonic branch and should have applications to the physics of flux compactifications with warped throats.

\end{abstract}

\section{Introduction}
The study of type II supergravity on the conifold has provided much insight into the low-energy limit of string theory. The explicit Ricci flat metric on the singular conifold, its small resolution and its deformation were all computed in \cite{Candelas:1989js} and immediately after the advent of the $AdS$-CFT correspondence  \cite{Maldacena:1997re,Gubser:1998bc,Witten:1998qj} it was realized that the singular conifold, suitably embedded in IIB string theory admits a near horizon limit that contains an $AdS_5$ factor, and describes the supergravity dual of a certain superconformal field theory \cite{Klebanov:1998hh}. This system and its generalizations have proved to be a fertile ground for the study of the gauge/gravity duality. In particular
the ``warped deformed conifold'' non-conformal deformation found in \cite{Klebanov:2000hb}  (which we will denote KS) is an explicit supergravity realization of confinement and chiral symmetry breaking. The solution found in \cite{Chamseddine:1997nm, Maldacena:2000yy} (which we will denote CV/MN) is the holographic dual of a field theory whose infra-red limit is four-dimensional $\N=1$ Yang-Mills theory. The rich physics of the above supergravity solutions led to the proposal of an anzatz (which we will call the PT ansatz)  \cite{Papadopoulos:2000gj} that describes the solutions interpolating between the KS and the CV/MN flows. Following the perturbative analysis of \cite{Gubser:2004qj}, a fully backreacted supersymmetric solution interpolating between the KS and the CV/MN solutions was constructed in \cite{Butti:2004pk} (since this solution is dual to the baryonic branch of the warped deformed conifold we will denote it BB). Recently an interesting conceptual development regarding this system was provided \cite{Maldacena:2009mw}.

The current work constitutes a step towards improving our understanding of the spectrum of both supersymmetric and nonsupersymmetric deformations around the warped deformed conifold and its baryonic branch (KS and BB). A thorough understanding of these deformations is important for many problems in string cosmology and phenomenology and there has been much work on computing the spectrum of fluctuations around KS  (for example \cite{Berg:2005pd, Berg:2006xy, Dymarsky2008,Benna:2007mb, Dymarsky:2008wd,Gordeli2009}) but to our knowledge certain $SU(2)\times SU(2)$ invariant modes which lie within our ansatz have not been considered. These modes are crucial for constructing a supersymmetric five-dimensional theory and thus are needed to arrange the spectrum into supermultiplets.

The PT ansatz was constructed so as to contain a variety of conifold solutions in IIB supergravity and indeed families of new solutions were found within this ansatz \cite{Butti:2004pk}. However this ansatz is deficient is several ways, firstly it does not contain one-form or two-form fields in five dimensions and secondly there are seven $SU(2)\times SU(2)$ invariant scalar modes which are absent. By systematically including all singlets under this symmetry we are able to construct an explicitly supersymmetric action in five dimensions. This has been an open problem for some time (it is ten years since the PT ansatz was written down and six years since explicit solutions were found) but recent progress in Kaluza-Klein reductions (see \cite{Gauntlett:2007ma, Gauntlett:2009zw} for early work and  \cite{Cassani:2010uw, Gauntlett:2010vu, Liu:2010sa, Skenderis:2010vz} for more recent additions) has demonstrated clearly that a promising strategy is at hand. These works were focused on finding infinite classes of reductions on manifolds using minimal information about the internal manifolds (namely the Sasaki-Einstein structure) whereas we will focus solely on $T^{1,1}$. See also \cite{Herzog:2009gd} for recent consistent (non-supersymmetric) truncations on $T^{1,1}$.

One central motivation for the current work is to compute a superpotential in five dimensions for a reduction which includes PT. Such a superpotential would provide an organizing principle for the spectrum of supersymmetric and non-supersymmetric modes \cite{Borokhov:2002fm}. Recently, several of the current authors \cite{Bena:2009xk} have computed a certain sector of the spectrum around the warped deformed conifold in search of the modes sourced by a stack of anti-D3 branes placed at the tip of the deformed conifold. Having a superpotential at hand for this sub-sector was a crucial step in the calculation and also considerable help in the physical interpretation. With this superpotential it is manifestly clear which modes are BPS and which modes are not. 

There exists a very interesting conjecture that a stack of anti-D3 branes at the tip of the warped deformed conifold will create a meta-stable vacuum and ultimately decay by brane-flux annihilation to the BPS vacuum \cite{Kachru:2002gs}. In the work \cite{Bena:2009xk} we have carefully imposed IR boundary conditions compatible with anti-D3 branes and have found a seemingly unphysical singularity. We have interpreted this singularity as evidence that the IR boundary conditions of anti-D3 branes are incompatible with the UV boundary conditions of KS. Since there must exist good IR boundary conditions we conclude that the UV boundary conditions must be changed. 
Establishing this conjecture to be true is probably tantamount to constructing the supergravity solution dual to this state explicitly. 
Then to demonstrate that such a candidate solution is in fact a metastable vacuum, it behooves one to check that indeed it does not have runaway directions. Rigorously, this would entail checking all modes  but this not a reasonable task in general. Having this consistent truncation may be helpful in this regard in checking at least a closed subsector should such a solution ever be constructed. However recently there has been an example found where a non-supersymmetric solution \cite{Pope:1984bd} is stable within a consistent truncation but has runaway directions which lie outside the truncation \cite{Bobev:2010ib}. This is a general limitation of consistent truncations.

This paper is organized as follows: In section two we present our ansatz and proceed to derive the five-dimensional action. Then in section three we demonstrate that this is consistent with $\N=4$ supersymmetry and compute the embedding tensor. In section four we consider various additional consistent truncations including the PT ansatz. The reduction techniques used in this paper are very similar to those used in \cite{Gauntlett:2009zw} and the manifest supersymmetry is demonstrated following closely the strategy of \cite{Cassani:2010uw, Gauntlett:2010vu}. In an attempt to minimize the amount of labor required for interpreting our work, we have tried to use notation quite similar to \cite{Gauntlett:2010vu} for the bulk of the computation. The same day we submitted this paper to the arxiv a paper appeared by D. Cassani and A. Faedo \cite{Cassani:2010na} which has some overlap with this work.

\section{The Kaluza-Klein Reduction on $T^{1,1}$}
Our Kaluza-Klein reduction on $T^{1,1}$ is best thought of as a consistent truncation of
the standard reduction of IIB on $T^5$ (where of course only massless modes are retained), followed by a very specific gauging which preserves $\N=4$ supersymmetry in five-dimensions. This gauging is due to the curvature of $T^{1,1}$ as well as the topological flux which we include. 

To perform this Kaluza-Klein reduction, we can simply dimensionally reduce many parts of the ten-dimensional action, namely the kinetic terms for the dilaton-axion, three-forms and metric but the five-form kinetic terms and the Chern-Simons terms require much more care. We will proceed by deriving five-dimensional equations of motion for components of the five-form and then reconstructing a five-dimensional action from these. Since our reduction is based on a symmetry principle, it is guaranteed to be consistent.

\subsection{The Ansatz}
We will motivate our full ansatz by first understanding the structures we wish to preserve on the metric of $T^{1,1}$. Then we will make an ansatz for the three-forms and five-form and solve the Bianchi identities. The dilaton and axion are additional fields which have trivial Bianchi identities.

\subsubsection{The Metric} \label{sec:metric}

Our ansatz for the ten-dimensional metric is
\bea
ds_{10}^2&=&  e^{2u_3-2u_1} ds_5^2 + e^{2u_1+2u_2} E'_1 \Ebar'_1 +e^{2u_1-2u_2} E'_2 \Ebar'_2+ e^{-6u_3-2u_1} E_5 E_5 \label{metansatz},
\eea
where
\bea
E_1=\frac{1}{\sqrt{6}}\blp  \sig_1+i\sig_2 \brp, &&E_2=\frac{1}{\sqrt{6}}\blp \Sig_1+i\Sig_2 \brp ,\\
E'_1=E_1 && E'_2= E_2+v \Ebar_1,\label{fr1}\\
E_5 = g_5+A_1, && g_5=\frac{1}{3}\blp \sig_3+\Sig_3\brp,
\eea
$(\sig_i,\Sig_j)$ are internal left-invariant $SU(2)$ one-forms  (see appendix \ref{app:conventions}), the scalar fields $u_j$ are real, $v$ is complex and $A_1$ is a real one-form, all in the reduced five-dimensional theory. The particular parameterization we have chosen for the scalar fields in \eq{metansatz} might seem convoluted but is motivated by getting canonical kinetic terms in the five-dimensional action.

This is the most general $SU(2)\times SU(2)$ invariant metric we can put on $T^{1,1}$ and consequently, by introducing a radial co-ordinate in five dimensions and allowing for the fields to have radially dependent profiles this ansatz contains the most general $SU(2)\times SU(2)$ invariant metric on the conifold (singular, resolved or deformed).

In terms of $G$-structures, this is a $U(1)$ structure on $T^{1,1}$, namely we have restricted to all 
$U(1)$ neutral fields where the $U(1)$ in question is embedded in $SO(5)$ as follows:
\be
U(1)\subset SU(2)_D \subset SO(4)\subset SO(5)
\ee
and $SU(2)_D$ is diagonally embedded in $SO(4)$. As can be seen from tables \ref{table:Gdecompsc} and \ref{table:Gdecompvec} in appendix \ref{app:Gstructures}, this leaves 16 scalars, 9 one-forms and 1 graviton. Under this $U(1)$ the vielbeins transform as 
\be
E_1\ra e^{i\al}E_1, \ \ \  E_2\ra e^{-i\al}E_2,\ \ \ E_5\ra E_5 \non 
\ee
and from this it is simple to observe that the invariant fundamental forms are
\bea
J_1=\frac{i}{2} E_1\w \Ebar_1, && J_2=\frac{i}{2} E_2\w \Ebar_2, \non \\
\Om=E_1\w E_2,&& \Ombar=\Ebar_1\w \Ebar_2, 
\eea
as well as $g_5$. From \eq{fr1} we see that these forms behave nicely under the exterior derivative
\bea
dg_5&=&2(J_1+J_2), \non \\
d J_{1,2} &=&0,  \label{diffrels}\\
d\Om &=&3i\Om\w g_5\non 
\eea
but it turns out to be prudent to introduce a twisted set of fundamental forms which behave nicely under the Hodge star operation (our conventions for Hodge dualizing are given in appendix \ref{app:hodge}). 
\bea
J'_1=\frac{i}{2} E'_1\w \Ebar'_1, && J'_2=\frac{i}{2} E'_2\w \Ebar'_2, \non \\
\Om'=E'_1\w E'_2,&& \Ombar'=\Ebar'_1\w \Ebar'_2 .
\eea
In fact this truncation can also be thought of as the $\cK$-invariant truncation where in standard co-ordinates on the conifold, $\cK$ is a particular $\ZZ_2$ symmetry acting on the conifold as 
\be
\cK: (\psi,\tha_2) \ra (\psi+\pi , -\tha_2).
\ee
Consequently, since our ansatz contains all modes invariant under this symmetry it must be consistent\footnote{See the introduction of \cite{Bobev:2010ib} for a review of the argument why a truncation to a sector invariant under a symmetry is consistent and a discussion of its limitations}.

\subsubsection{The Three-Forms}

The three-forms in our ansatz are
\bea
H_{(3)}&=& H_3 + H_2 \w (g_5+A_1) + H_{11} \w J_1 + H_{12}\w J_2 + \non \\
&& +   \Blp M_1\w \Om +M_0\, \Om  \w (g_5+A_1)+c.c \Brp,  \\
F_{(3)}&=& P (J_1-J_2)\w (g_5+A_1) + G_3 + G_2 \w (g_5+A_1) + G_{11} \w J_1 + G_{12}\w J_2 + \non \\
&& +   \Blp N_1\w \Om+ N_0\, \Om \w (g_5+A_1)+c.c  \Brp 
\eea
where we have included a topological term in $F_{(3)}$
\be
P (J_1-J_2)\w (g_5+A_1) 
\ee
which is proportional to the volume form on the topologically nontrivial $S^3\subset T^{1,1}$. One can also include an independent  topological term for the NS flux\footnote{In a forthcoming publication we will analyze more formal geometric and non-geometric properties of the general embedding tensor} but by using the IIB $SL(2,\ZZ)$ symmetry, this can always be rotated to a frame where the charge is just in $F_{(3)}$. 

To establish the spectrum from our ansatz we must first solve the Bianchi identities. From
\be
dH_{(3)}=0
\ee
we find
\bea
H_3&=& dB_2 +\half (db-2B_1) \w F_2, \non \\
H_2&=& dB_1, \non \\
H_{11}&=& d (b+\tb)-2B_1, \label{H3Bianchi} \\
H_{12}&=& d (b-\tb)-2B_1, \non\\
3i M_1&=& DM_0 \non \\
&=&dM_0 - 3i A_1 M_0\non
\eea
where $F_2=dA_1$. Then from
\be
dF_{(3)}= -F_{(1)}\w H_{(3)}
\ee
we find
\bea
G_3&=& dC_2-a\,dB_2 +\half (dc-adb-2C_1+2aB_1) \w F_2, \non\\
G_2&=& dC_1-a\,dB_1, \non \\
G_{11}&=& d (c+\tc)-2C_1 -a \blp d(b+\tb) -2B_1\brp -PA_1,   \label{F3Bianchi}\\
G_{12}&=& d  (c-\tc)-2C_1 -a \blp d(b-\tb) -2B_1 \brp+ PA_1, \non\\
3iN_1 &=& DN_0 + M_0 da \non \\
&=& dN_0 -3i A_1 N_0+ M_0 da \non
\eea
and $a$ is the RR axion $C_{(0)}$.
From these relations we discover that the fluxes contribute 8 scalars
\be
(b,\tb,c,\tc,M_0,\Mbar_0,N_0,\Nbar_0)
\ee
a pair of two-form potentials
\be
(B_2,C_2)
\ee
and two one-form potentials
\be
(B_1,C_1).
\ee
Using these fields one can of course write the three-form field strengths in terms of two-form potentials:
\bea
H_{(3)}&=& dB_{(2)}, \non \\
\Rightarrow B_{(2)} &=& B_2 +\half bF_2+ B_1\w(g_5+A_1) + (b+\tb)J_1 + (b-\tb)J_2 + ( \frac{1}{3i}M_0 \Om+c.c.), \\
F_{(3)}&=& P(J_1-J_2)\w (g_5+A_1) + dC_{(2)}-a\,dB_{(2)}, \non \\
\Rightarrow  C_{(2)}&=& C_2 +\half cF_2+ C_1\w(g_5+A_1) + (c+\tc)J_1 + (c-\tc)J_2 + (\frac{1}{3i} N_0 \Om+c.c).  
\eea

\subsubsection{The Five-Form}
We take the five-form to be manifestly self-dual 
\bea
F_{(5)}&=&e^Z e^{8(u_3-u_1)} \vol_5 + e^Z J'_1\w J'_2\w (g_5+A_1)  \non \\
&&+K'_1\w J'_1 \w J'_2  -e^{-8 u_1} (*_5 K'_1) \w (g_5+A_1) \non \\
&& + K'_{21}\w J'_1 \w (g_5+A_1) + e^{-4u_2+4u_3}(*_5K'_{21})\w J'_2 \non \\
&& + K'_{22}\w J'_2 \w (g_5+A_1) +e^{4u_2+4u_3} (*_5K'_{22})\w J'_1 \non \\
\myvspace
&& + \blp L'_{2} \w \Om'+c.c \brp \w (g_5+A_1 ) +  e^{4u_3} \blp (*_5L'_2)\w\Om' +c.c \brp\, ,  \label{fiveansatz}
\eea
where we have defined the primed forms such as $K'_1$ in Appendix \ref{app:fiveform}.

Due to this self-duality of the five-form, the Bianchi identity for the five-form must be disentangled from the equation of motion 
\be
dF_{(5)} = H_{(3)}\w F_{(3)} \non
\ee
and we find (see appendix \ref{app:fiveform})
\bea
e^Z&=& Q-2P \tb +\frac{4i}{3} \blp \Mbar_0 N_0 - M_0 \Nbar_0 \brp \label{F5Bianchi1} \\
K_1&=& Dk+2(bDc-\tb D\tc) + \frac{2i}{3}\Blp \Mbar_0 N_1-\Nbar_0 M_1 -M_0\Nbar_1 +N_0 \Mbar_1\Brp \label{myK1}\\
K_{21}&=& D k_{11} + \half \Bslb Db \w D c +D b\w D\tc + D \tb\w D c\Bsrb   \\
K_{22}
&=& D k_{12} + \half \Bslb Db \w D c -D b\w D\tc - D \tb\w D c\Bsrb   \label{F5Bianchi4}
\eea
where
\bea
Dk&=&dk+4 c B_1 -QA_1-2 (k_{11}+k_{12}), \non \\
D k_{11} = d k_{11} + P B_2, && D k_{12} = dk_{12} - PB_2, \label{F5Bianchi}\\
D b= db -2B_1,&& D\tb =d \tb,\non \\
D c= dc -C_1,&& D\tc =d \tc - PA_1 \non 
\eea
and $Q$ is a constant corresponding to the D3 Page charge.
So from the five-form we get:
\bea
{\rm scalar}&-&  k \non \\
{\rm 1-form}&-& (k_{11},k_{12}) \\
{\rm 2-form}&-& (L_2,\Lbar_2)\non
\eea 
Deriving the  equations of motion for the five-form is one of the more involved steps in the analysis, the results are presented in appendix \ref{app:fiveform}. The equations of motion are of course necessary to construct the five-dimensional Lagrangian, to which we now turn.

\subsection{The Five-Dimensional Lagrangian}

There are several subtleties in producing a five-dimensional Lagrangian whose equations 
of motion match those of the ten dimensional theory, largely due to 
the Chern-Simons terms in ten dimensions. We have checked that the Lagrangian we present below
indeed reproduces the ten dimensional equations of motion with the most non-trivial
task being to check the flux equations of motion.

The five-dimensional action is the sum of several terms 
\bea
\cL=\cL_{gr}+\cL_{s,kin}+\cL_{g,kin}+\cL_{pot}+\cL_{CS} \label{FullLag}
\eea
which correspond to the five-dimensional Einstein-Hilbert term, the scalar kinetic terms, the kinetic terms for the gauge fields and two-forms, the scalar potential and the Chern-Simons terms. We find the scalar kinetic terms to be
\bea
\cL_{s,kin}&=& -\half e^{-4(u_1+u_2)-\phi}H'_{11}\w * H'_{11} -\half e^{-4(u_1-u_2)-\phi}H_{12}\w * H_{12}- 4 e^{-4u_1-\phi}M'_1\w *\Mbar'_1  \non \\
&&-\half e^{-4(u_1+u_2)+\phi}G'_{11}\w * G'_{11} -\half e^{-4(u_1-u_2)+\phi}G_{12}\w * G_{12} - 4 e^{-4u_1+\phi}N_1'\w *\Nbar'_1 \non \\
&&-8d u_1\w*d u_1  -4d u_2\w*d u_2 -12 d u_3\w*d u_3 - e^{-4u_2} Dv \w * D\vbar \non \\
&&-\half e^{-8u_1} K_1\w * K_1  -\half d\phi \w * d\phi -\half e^{2\phi} da \w *da \label{Lskin},
\eea
where we have twisted some of the one-forms
\bea
H'_{11}&=&H_{11}-|v|^2 H_{12}- 4\, \Im \!(v M_1), \\
M'_1&=&M_1+\frac{i}{2} \vbar H_{12},  \\
G'_{11}&=&G_{11}-|v|^2 G_{12}- 4\, \Im \!(v N_1), \\
N'_1&=&N_1+\frac{i}{2} \vbar G_{12} 
\eea
and $\phi$ is the dilaton.
The kinetic terms for the gauge fields are
\bea
\cL_{g,kin}&=&-\half e^{-8u_3} F_2\w * F_2-\half e^{4u_1-4u_3-\phi} H_3\w * H_3- \half e^{4u_1+4u_3 - \phi} H_2\w *H_2 \non \\
&& -\half e^{4 u_1-4 u_3+\phi} G_3\w * G_3- \half e^{4u_1+4u_3+\phi} G_2\w *G_2  \non \\
&&-4 e^{4u_3}\blp 1+|v|^2e^{-4u_2}\brp L_2\w * \Lbar_2   + 4 e^{-4u_2+4u_3}\Blp v^2 L_2\w * L_2 + c.c \Brp \non \\
&& - \half e^{4u_2+4u_3}  \blp 1+|v|^2 e^{-4u_2} \brp^2K_{22}\w * K_{22} - \half e^{-4u_2+4u_3} K_{21}\w * K_{21} \non \\
&& +|v|^2 e^{-4u_2+4u_3}  K_{22}\w *K_{21}  + 2e^{4u_3} \blp 1+ |v|^2 e^{-4u_2}   \brp  \Blp i v K_{22}\w *L_2 +c.c \Brp\non \\
&& -2e^{-4u_2+4u_3}  \Blp iv K_{21}\w *L_2 +c.c \Brp
\eea
where the somewhat off-diagonal last four lines come from the five-form.

The scalar potential has several contributions which we distinguish for clarity:
\bea
\cL_{pot}&=&-\blp V_{gr}+  V_{H_{(3)}}+V_{F_{(3)}}+V_{F_{(5)}}\brp, \label{potential}\\
V_{gr}&=&  -12e^{-4u_1-2u_2+2u_3}\blp 1+|v|^2 +e^{4u_2}\brp  +9|v|^2e^{-4u_2+8u_3}  \non \\
&&+2e^{-8u_1-4u_3} \blp e^{4u_2}+e^{-4u_2}(1-|v|^2)^2+2|v|^2\brp, \\
V_{H_{(3)}}&=&4 e^{-4u_1+8u_3-\phi}\Blp  |M_0|^2  +2e^{-4u_2}\slb\Im\!(M_0 v)\srb  ^2 \Brp, \\
V_{F_{(3)}}&=&\half e^{-4u_1+8u_3+\phi} \Blp 8 |N'_0|^2  +e^{4u_2} P^2+e^{-4u_2}  \blp P(|v|^2-1)+4\, \Im\!(N'_0 v) \brp^2\Brp,   \\
V_{F_{(5)}}&=& \half e^{2Z} e^{-8u_1+8u_3}
\eea
where
\be
N'_0=N_0-\frac{i}{2} P \vbar.
\ee
As expected this scalar potential is almost but not quite a sum of squares. The only term which spoils this property is $V_{gr}$. Finally the gravitational term is of course
\be
\cL_{gr}=R\, \vol_5
\ee
where $R$ is the Ricci scalar in Einstein frame. The Chern-Simons terms are particularly long and unspectacular so we will not write them explicitly. In the ungauged case which we deal with below, they are somewhat simpler and also extremely crucial so we will present them explicitly there.

\section{Manifest $\N=4$ Supersymmetry}
A particularly insightful aspect of the works \cite{Cassani:2010uw, Gauntlett:2010vu} was the construction of manifest $\N=4$ supersymmetry (by which we mean 16 supercharges). In that case, the reason this was unexpected was that this particular gauging of $\N=4$ supergravity does not have a vacuum which preserves all the supercharges, the maximally supersymmetric vacuum is an $AdS_5$ which preserves only $\N=2$. Still more surprisingly, despite the fact that we have generalized the reduction considered there by including 5 new scalars and one new vector, the $\N=4$ supersymmetry is still present in our work. We find that these extra fields constitute just one additional $\N=4$ vector multiplet compared to those works.

\subsection{The scalar coset}
In five dimensional, $\cN=4$ supergravity there are the following multiplets
\bitem
\item graviton multiplet: $(g_{\mu\nu},6\times A_\mu, \phi) $
\item vector multiplet: $(A_\mu, 5\times \phi)$
\eitem
so clearly our reduction has the bosonic field content of $\N=4$ gauged supergravity coupled to three vector multiplets\footnote{The half maximal  gauged supergravity theories in four and five dimensions are quite rigid and are summarized nicely in \cite{Dall'Agata:2001vb, Schon:2006kz}. Another very useful source of information is the nice lecture notes \cite{Samtleben:2008pe}}. The scalar coset of this five-dimensional $\N=4$ gauged supergravity is \cite{Awada:1985ep}
\be
\frac{SO(5,3)}{SO(5)\times SO(3)} \times SO(1,1)
\ee
and a particular basis for this coset was given in \cite{Lu:1998xt} eq. (3.31). In the conventions of that paper we take our coset element to be
\bea
\cV&=&e^{2^{3/2}\phi_1 H_1}e^{2^{-3/2}\phi_2 H_2}e^{2^{-3/2}\phi_3 H_3}e^{x_1 E_2^{3}}e^{x_2 E_1^{3}}e^{x_3 E_1^{2}} \non \\
&&.e^{x_7 U_1^{1}}e^{x_8 U_1^{2}}e^{x_9 U_1^{3}}e^{x_{10} U_2^{1}}e^{x_{11} U_2^{2}}e^{x_{12} U_2^{3}} \non \\
&& .e^{x_4 V^{23}}e^{x_5 V^{13}}e^{x_6 V^{12}}  \label{cosetelement}
\eea
and the scalar Kinetic terms are
\be
\cL_{s,kin}= -3\Sig^{-2}d\Sig\w * d\Sig + \frac{1}{8}\Tr (dM \w *dM^{-1}) 
\ee
where
\be
M=\Tr \cV \cV^T.
\ee
Explicitly we find
\bea
-\Tr (dM \w *dM^{-1})&=& 2 \blp d\phi_1^2 + d\phi_2^2 + d\phi_3^2\brp \non \\
&&+4e^{-\phi_2+\phi_3} dx_1^2 +4e^{-\phi_1+\phi_3} (dx_2-x_1 dx_3)^2 +4e^{-\phi_1+\phi_2} dx_3^2\non \\
&&+4e^{\phi_1+\phi_2} (dx_4+x_7 dx_8+x_{10}dx_{11})^2 \non \\
&&+4e^{\phi_1+\phi_3} \blp dx_5+x_7 dx_9+x_{10}dx_{12}-x_1(dx_4+x_7 dx_8+x_{10}dx_{11})\brp^2 \non \\
&&+4e^{\phi_1}dx_7^2 +e^{\phi_2} \blp dx_8-x_3 dx_7 \brp^2\non \\
&&+4e^{\phi_3} \blp dx_9-(x_2-x_1x_3) dx_7 -x_1 dx_8\brp^2 \non \\
&&+4e^{\phi_1} dx_{10}^2+e^{\phi_2} \blp dx_{11}-x_3 dx_{10} \brp^2\non \\
&&+4e^{\phi_3} \blp dx_{12}-(x_2-x_1x_3)  dx_{10} -x_1 dx_{11}\brp^2 \non \\
&&+ 4e^{\phi_2+\phi_3} \bslb dx_6+x_2dx_4 +x_2x_7 dx_8+x_2 x_{10}dx_{11}- x_3 dx_5   \non \\
&&+(x_8-x_3 x_7)dx_9 +(x_{11}-x_3 x_{10})dx_{12} \bsrb^2 . \label{gpmet}
\eea
It may be helpful to describe how this basis (which we will refer to as the ``heterotic basis") is related to a more common basis in the gauged supergravity literature \cite{Schon:2006kz,Samtleben:2008pe}  (which we will refer to as the ``gsg basis") where generators of $SO(5,3)$ are given by
\be
(t_{MN})_{P}^{\ \ Q}= \delta^Q_{[M}\eta_{N]P} \label{basistMN}
\ee
$M,N\ldots = 1,\ldots,8$ and $\eta={\rm diag}\{ +++++---\}$. Of course only a subset of the $t_{MN}$ generate the coset $SO(5,3)/(SO(5)\times SO(3))$. The two basis (where the Heterotic basis is completed to a full set of generators of $SO(5,3)$) are related by conjugation with $C$:
\be
C= D_1 + D_2 +D_3 + E_{44}+ E_{55},
\ee
where
\be
D_i=\blp  E_{i,i}-E_{i,i+5}+E_{i+5,i}+E_{i+5,i+5}\brp /\sqrt{2} \ \ \ i=1,\ldots,3
\ee
and where $E_{ij}$ is a matrix with $1$ in the $i$-th row and $j$-th column and zero's elsewhere.  This is most easily seen by relating $\eta$ and $\teta$ where
\be
\barr{rl}
\teta_{16}=&\teta_{61}=-1 \\
\teta_{27}=&\teta_{72}=-1 \\
\teta_{38}=&\teta_{83}=-1 \\
\teta_{44}=&\teta_{55}=1
\earr \label{etamat}
\ee
and $\cV^T \teta \cV= \teta$. 

Matching \eq{gpmet} to our scalar kinetic terms \eq{Lskin} obtained from dimensional reduction we find the rather nonlinear (but  invertible) mapping
\bea 
e^{2u_3}&=& \Sig \non \\
 -4u_2&=& \phi_1 \non \\
 -4u_1 - \phi &=& \phi_2 \non \\
 - 4 u_1 + \phi &=&\phi_3\non \\
 \sqrt{2}\,v&=&x_7+ix_{10} \non \\
a&=&x_1 \non \\
b-\tb&=&x_3  \non \\
b+\tb &=& -x_4-\half x_3 (x_7^2+x_{10}^2) \non \\
c-\tc &=& x_2\label{scalardefs} \\
c+\tc&=& -x_5 -\half  x_2(x_7^2+x_{10}^2) \non\\
\frac{2\sqrt{2}}{3}M_{0}&=&- \blp x_8-x_3 x_7\brp  +i (x_{11}-x_3x_{10})\non  \\
&=& -(x_8-ix_{11}) + x_3 (x_7-ix_{10}) \non \\
\frac{2\sqrt{2}}{3} N_0 &=&-\blp x_9 - (x_2-x_1 x_3) x_7 - x_1 x_8\brp + i\blp x_{12}- (x_2-x_1x_3) x_{10} -x_1 x_{11} \brp  \non \\  
&=& -(x_9-ix_{12})+ (x_2-x_1x_3) (x_7-i x_{10}) +x_1 (x_8-ix_{11}) \non \\
k&=& x_6+ x_2 x_4 +\half x_2x_3 (x_7^2+x_{10}^2)\non \\
&&+\half \Blp x_2\blp x_7x_8+x_{10}x_{11}\brp + x_9(x_8-x_3x_7)+x_{12}(x_{11}-x_3 x_{10})\Brp. \non
\eea

\subsection{The gauge kinetic and Chern-Simons terms in the ungauged reduction}
In ungauged $\N=4$ supergravity, the kinetic and Chern-Simons terms for the gauge fields take a particularly simple form:
\be
\cL_{g,kin}+\cL_{CS}=-\half\Sig^{-4} \cH_2^0\w *\cH^0_2 -\half \Sig^{2} M_{MN} \cH_2^M \w *\cH_2^N + \frac{1}{\sqrt{2}} \eta_{MN} {\cal A}_1^0 \w \cH_2^M \w \cH_2^N \label{lagformal}
\ee
where $\cH_2^M = \cD \cA^M$ are field strengths for the gauge fields. Indeed this is what we find by setting the topological fluxes $(P,Q)$ to zero and altering the differential relations \eq{diffrels} to
\be
dg_5=0,\ \ \ dJ_{1,2}=0,\ \ \ d\Om=0.
\ee
With these alterations, massaging the Lagrangian \eq{FullLag} into this form allows one to identify the correct basis to take for the nine gauge fields, however first we must integrate out the pair of two-forms $(B_2,C_2)$. The central difference between the gauged reduction and the ungauged reduction is the Bianchi identities and their solution: instead of \eq{H3Bianchi}, \eq{F3Bianchi} and \eq{F5Bianchi1}-\eq{F5Bianchi4} we have
for $H_{(3)}$:
\be
\begin{array}{ll}
H_3=dB_2 -B_1\w F_2, \ \ \ \ \ \ \ & H_2 = dB_1, \\
H_{11}= d(b+\tb),  & H_{12}= d(b-\tb), \\
3i M_1= dM_0, &  
\end{array}
\ee
for $F_{(3)}$
\be
\begin{array}{ll}
G_3= dC_2-adB_2 -(C_1-aB_1) \w F_2,\ \ \ \ \ \ \ \ 
& G_2= dC_1-a\,dB_1, \\
G_{11}= d (c+\tc) -a\, d(b+\tb), 
& G_{12}= d  (c-\tc) -a \, d(b-\tb), \\
3iN_1=dN_0-adM_0, &
\end{array}
\ee
and for $F_{(5)}$
\be
\barr{rl}
K_1=& dk+2(bdc-\tb d\tc)  + \frac{2i}{3}\Blp \Mbar_0 N_1-\Nbar_0 M_1 -M_0\Nbar_1 +N_0 \Mbar_1\Brp, \\
K_{21}=& dk_{11} + (b+\tb)dC_1 -(c+\tc)dB_1, \\
K_{22}=& dk_{12} + (b-\tb)dC_1 -(c-\tc)dB_1,  \\
L_2=& dD_1 +\frac{1}{3i} \blp M_0 dC_1 - N_0 dB_1\brp  .
\earr
\ee

We find that before integrating out $(B_2,C_2)$, the Chern-Simons terms are
\bea
\cL_{top}&=&-A_1  \w \Bslb  K_{22}\w K_{21}+ K_1\w (-C_1\w H_2+ B_1\w  G_2) + 4L_2\w \Lbar_2  \non \\
&&+  K_{21}\w \bslb d(b-\tb)\w C_1 - d(c-\tc)\w B_1\bsrb  \non \\
&&+ K_{22}\w \bslb d(b+\tb)\w C_1 - d(c+\tc)\w B_1\bsrb \non \\
&& +\blp (4i/3)L_2\w (\Mbar_0 C_1 -\Nbar_0B_1)+c.c\brp   \Bsrb \non \\
&& - dC_2 \w S_2 + dB_2 \w T_2 
\eea
where
\bea
S_2 &=& \blp k +\frac{4}{9}\Re\!\! (M_0 \Nbar_0)\brp dB_1 -(b^2 - \tb^2 + \frac{1}{9} |M_0|^2) dC_1\non \\
&& - (b-\tb) d k_{11}-(b+\tb) d k_{12} -\frac{8}{3} \Im\! \blp M_0 d\Dbar_1\brp \\
T_2 &=& \blp k-\frac{4}{9}\Re\!\! (M_0 \Nbar_0)\brp dC_1+(c^2 - \tc^2 + \frac{1}{9} |N_0|^2) dB_1\non \\
&&  - (c-\tc) d k_{11} -(c+\tc) d k_{12} - \frac{8}{3}\Im\!\blp N_0 d\Dbar_1\brp.
\eea
First we introduce Lagrange multiplers $(\widetilde{B}_1,\tC_1)$
\be
\Delta \cL = \tC_1 \w d\tH_3 + \tB_1 \w d\tG_3
\ee
where 
\be
 \tH_3= dB_2,\ \ \  \tG_3=dC_2, \non \\
\ee
and then we integrate out $(\tH_3,\tG_3)$ and after some algebra we find 
\bea
\cL_{g,kin}&=&-\half e^{-8u_3} F_2\w * F_2-\half e^{-4(u_1-u_3)-\phi} \tH_2\w * \tH_2-\half e^{-4(u_1-u_3)+\phi} \tG_2\w * \tG_2 \non \\
&& - \half e^{4(u_1+u_3)-\phi} H_2\w *H_2- \half e^{4(u_1+u_3)+\phi} G_2\w *G_2  \non \\
&&-4 e^{4u_3}\blp 1+|v|^2e^{-4u_2}\brp L_2\w * \Lbar_2   + 4 e^{-4u_2+4u_3}\Blp v^2 L_2\w * L_2 + c.c \Brp \non \\
&& - \half e^{4u_2+4u_3}  \blp 1+|v|^2 e^{-4u_2} \brp^2K_{22}\w * K_{22} - \half e^{-4u_2+4u_3} K_{21}\w * K_{21} \non \\
&& + |v|^2 e^{-4u_2+4u_3}  K_{22}\w *K_{21}  + 2e^{4u_3} \blp 1+ |v|^2 e^{-4u_2}   \brp  \Blp i v K_{22}\w *L_2 +c.c \Brp\non \\
&& -2e^{-4u_2+4u_3}  \Blp iv K_{21}\w *L_2 +c.c \Brp, \label{Lgking2}\\
\cL_{top}&=&  -A_1 \w \Bslb dk_{12}\w dk_{11}  +4dD_1\w d\Dbar_1 - dB_1 \w d\tC_1 -dC_1\w d\tB_1\Bsrb, \label{Ltop2} 
\eea
Identifying (\ref{Lgking2}) with \eq{lagformal} we construct the $SO(5,3)$ vector of one-form potentials:
\be
\barr{ll}
\cA^0= -A_1/\sqrt{2}, & \\
\cA^1 = -k_{11}/\sqrt{2},\ \ \ \ \  &\cA^2 = \tB_1/\sqrt{2}, \\
\cA^3 = \tC_1/\sqrt{2},\ \ \ \ \ & \cA^4 =2 \Im\!(D_1), \\
\cA^5 =2 \Re\! (D_1), & \cA^6 =k_{12}/\sqrt{2}, \\
\cA^7 =  C_1/\sqrt{2}, &\cA^8 = B_1/\sqrt{2} .
\earr \label{vecbasis}
\ee

So we have now shown that the ungauged theory, corresponding to a particular consistent truncation on the five-torus $T^5$ has $\N=4$ supersymmetry. Of course this is not the theory of most interest to us but was a necessary step in developing the gauged theory, which corresponds to a consistent truncation on $T^{1,1}$.

\subsection{The embedding tensor for the gauged theory}
Having successfully demonstrated the manifest supersymmetry of  the ungauged theory, 
the content of the gauged theory can be neatly summarized in the embedding tensor. 
For our purposes, the embedding tensor has components  $(f_{MNP},\xi_{MN})$
which can be computed from the scalar kinetic terms:
\be
\cD M_{MN}=d M_{MN} + 2 \cA^P f_{P(M}^{\ \ \ \ Q}M_{N)Q} +2 \cA^0 \xi_{(M}^{\ \ P} M_{N)P}. \label{covder}
\ee

We have found the  heterotic basis \eq{cosetelement} to be computationally efficient but the 
embedding tensor is most naturally expressed in the ``gsg basis" \eq{basistMN} where it is completely antisymmetric (with all indices lowered):
\be
f_{MNP}=f_{[MNP]},\ \ \ \ \xi_{MN}=\xi_{[MN]}.
\ee
Note that our expressions \eq{vecbasis} are in the heterotic basis.
After some work explicitly computing \eq{covder}, we find that the non-vanishing components in the basis \eq{basistMN} are
\bea
&&f_{123}=-f_{128}=f_{137}=f_{178}=2, \\
&& \xi_{23}=-\xi_{28}=\xi_{37}=\xi_{78}=-Q/\sqrt{2}, \\
&& \xi_{45}=-3\sqrt{2} \\
&& \xi_{36}=\xi_{68}=\sqrt{2}P \label{xiP}
\eea
and permutations.

From this one can read off the covariant field strength of $\N=4$ gauged supergravity:
\be
\cH^M= d\cA^M + \half f_{NP}^{\ \ \ M}\cA^N\w \cA^P + \half \xi_{P}^{\ \ M}\cA^0\w A^P
\ee
and for example we have
\bea
\cH^1+ \cH^6 &=& d (\cA^1+\cA^6)-2\sqrt{2} \cA^7\w \cA^8, \non \\
\cH^7&=& d \cA^7,  \\
\cH^8&=& d \cA^8 \non 
\eea
here we have used the basis of gauged fields in the ``heterotic basis" \eq{vecbasis}.
From this we see the same Heisenberg algebra which was observed in \cite{Cassani:2010uw, Gauntlett:2010vu}. The only additional gaugings in our ansatz arise from the topological flux we have turned on \eq{xiP}. So the additional vector multiplet we have included has enhanced the complexity of the embedding tensor somewhat indirectly through the additional degrees of freedom required to allow for non-trivial topology and thus the flux $P$.

\subsection{The Scalar Potential}
A useful check of our computations is to compute the scalar potential from the gauged supergravity formula
\bea
V&=&\frac{1}{2}f_{MNP} f_{QRS} \Sigma^{-2}\Blp \frac{1}{12}M^{MQ} M^{NR}  M^{PS} - \frac{1}{4} M^{MQ}\eta^{NR} \eta^{PS} + \frac{1}{6} \eta^{MQ} \eta^{NR} \eta^{PS}\Brp \non \\
&& + \frac{1}{8} \xi_{MN} \xi_{PQ} \Sigma^4 \Blp M^{MP}M^{NQ}-\eta^{MP} \eta^{NQ} \Brp 
 + \frac{1}{6} \sqrt{2} f_{MNP} \xi_{QR} \Sigma M^{MNPQR} . \label{Vgsg}
\eea
where
\be
M_{MNPQR}= \eps_{abcde}\cV_M^{\ a}\cV_N^{\ b}\cV_P^{\ c}\cV_Q^{\ d}\cV_R^{\ e}.
\ee
$a=1,\dots,5$ are $SO(5)$ indices and $\cV$ is the coset element \eq{cosetelement}. To work with the $SO(5)$ indices it is best to transform to the ``gsg" basis for the coset. Note that the three separate terms in this expression are distinguished by the power of $\Sigma=e^{2u_3}$ and each such term is easily identified in \eq{potential}. After some calculation, we find agreement between the two expressions.

\section{Further Consistent Truncations}
\subsection{The PT Ansatz}
As mentioned in the introduction, one central motivation for the current work is to understand 
in detail the Kaluza-Klein reduction employed in \cite{Papadopoulos:2000gj}. Since we have the most general $SU(2)\times SU(2)$ invariant reduction, it is guaranteed that the reduction of  \cite{Papadopoulos:2000gj} lies within ours but the former  has no vector fields and so clearly cannot be supersymmetric. In the scalar sector it is obtained from our truncation by setting the following fields to zero
\be
(a,\Im\! M_0,\Re\! N_0, c,\tc,k, \Im\! v)\ra 0
\ee
leaving nine scalars. Although it is conceivable that this scalar sector alone could be supersymmetrized in some other way, we in fact find that the supersymmetrization of this ansatz requires further scalar fields to be included. 

We now demonstrate that indeed the PT truncation is consistent. Partial results in this direction were presented in the nice work \cite{Berg:2005pd}. The most striking feature of the PT ansatz is that there is a distinct asymmetry between the RR and NS three-forms. One can understand this as a direct consequence of setting the axion to zero and satisfying the equation of motion for the axion \eq{eom1}. Following this logic explicitly in our new truncation,  requires as a first step, certain choices in setting the source for the axion to zero:
\bea
-d(*e^{2\phi}da)&=& e^{4(u_1-u_3)} H_3\w *G_3 +e^{-4(u_1-u_3)} H_2\w *G_2 \non \\
&& +e^{-4(u_1+u_2)} H'_{11}\w *G'_{11}  +e^{-4(u_1-u_2)} H_{12}\w *G_{12}  \non \\
&&+4e^{-4u_1} \blp M'_1\w * \Nbar'_1 +c.c. \brp +\Blp 4e^{-4u_1+8u_3}\blp M_0\Nbar'_0+c.c. \brp \non \\
&&-4e^{-4u_1-4u_2+8u_3}\,\Im\!\!(M_0v)\blp P(1- |v|^2)-4\Im\!\!(N'_0v)\brp\Brp\vol_5 .
\eea
We achieve this by setting 
\be
(\Im M_0,\Re N_0,\Im\! v,c,\tc)\ra 0
\ee
along with
\be
(C_2,C_1)\ra 0.
\ee
Further inspection of the equations of motion which arise from the ten-dimensional three form equations of motion \eq{eom2} and \eq{eom3} reveal that in addition we must have 
\be
(k,B_2,B_1,k_{11},k_{12},L_2,\Lbar_2)\ra 0.
\ee
From our formalism these steps are quite straightforward but one equation requires additional work. 
Since we have set $B_1\ra 0$ we must set the various source terms in \eq{H2eom} to zero and this would appear to give a differential constraint amongst several of the remaining scalars $(b,\tb,\Re \!v, \Re \! M_0)$:
 \bea
0&=&H_{11} (1-|v|^2)e^{-4u_2}  + H_{12}\Blp e^{4u_2} -|v|^2(1-|v|^2)e^{-4u_2} +2|v|^2  \Brp \non \\
&& + 4\,\Im\! (vM_1) \Blp 1-(1-|v|^2)e^{-4u_2}    \Brp.\label{Hconstr1}
\eea
However if we take the exterior derivative of the Hodge star of \eq{Hconstr1} 
we in fact recover a linear combination of  \eq{H12eom},\eq{H11eom},\eq{M1eom}. This is the only non-trivial step in showing that the ansatz employed in \cite{Papadopoulos:2000gj} is indeed a consistent truncation.
 
Since it has appeared quite straightforward within our $\N=4$ truncation to show the consistency of the PT ansatz, it is natural to wonder if there is an internal symmetry at work and we now follow a brief digression with regards to this idea. Such a symmetry must differentiate between the NS and RR three-forms and as such the best candidate is worldsheet parity reversal under which:
\be
\Om_p:(g,\phi,B_{(2)},C_{(0)},C_{(2)},C_{(4)})\ra (g,\phi,-B_{(2)},-C_{(0)},C_{(2)},-C_{(4)})\non.
\ee
A geometric symmetry is also needed and the best candidate appears to be reversal of all internal co-ordinates:
\bea
\sig: (\sig_1,\Sig_1,g_5) &\ra& -(\sig_1,\Sig_1,g_5),  \non \\
 (\sig_2,\Sig_2) &\ra& (\sig_2,\Sig_2),  \non
 \eea
 which translates to 
 \bea
 \sig: (g_5,J_1,J_2,\Im\! \Om)&\ra& - (g_5,J_1,J_2,\Im\! \Om), \non \\
 \Re\! \Om &\ra & \Re\! \Om \non.
 \eea
At this point we see that $\vol_5$ and $\vol_{T^{1,1}}$ transform with opposite sign:
\bea
\Om_p\cdot \sig: (\vol_5,\vol_{T^{1,1}})\ra (-\vol_5,\vol_{T^{1,1}})
\eea
so we are forced to append five-dimensional parity $P_5$ and finally
\be
\cJ=\Om_p\cdot \sig \cdot P_5
\ee
appears to be our best candidate for a symmetry principle which restricts one to the PT ansatz. However $\cJ$ does not commute or anti-commute with the exterior derivative, to be more precise it 
anti-commutes with the external exterior derivative but commutes with the internal one. This means that  
even if terms in the potential have equal $\cJ$-charge, the field strength will not. As a result we conclude that there is no symmetry principle which restricts one to the PT ansatz, this is of course not a disaster at all because in the current work we have provided an explicitly supersymmetric embedding of the PT ansatz into a consistent truncation which {\it is} based on a symmetry principle.

\subsection{The $\cI$ Truncation} 
 Another interesting truncation is the restriction to modes which are invariant under the so-called $\cI$ symmetry:
 \bea
 &&\cI= \Om_p\cdot (-1)^{F_L}\cdot \sigma, \non \\
&& \sigma: (\tha_1,\phi_1,\tha_2,\phi_2)\ra(\tha_2,\phi_2,\tha_1,\phi_1) .
 \eea
This is the $\ZZ_2$ symmetry which is broken away from the origin of the baryonic branch in the Klebanov-Strassler gauge theory \cite{Gubser:2004qj}. This truncation of our $\N=4$ reduction eliminates $(b,c)$ from the three form fluxes and the metric mode which is the coefficient of $E_1\Ebar_1- E_2\Ebar_2$. In addition only two vector fields remain, namely $A_1$ and the combination $k_{11}+k_{12}$. It seems plausible that this is the bosonic content of an $\N=2$ gauged supergravity in five dimensions with one vector multiplet and three hypermultiplets, it would be interesting to develop this further.

\section{Conclusions} 

In this work we have constructed a five-dimensional gauged supergravity theory 
by explicit dimensional reduction on $T^{1,1}$, including the entire set of modes which are singlets under the global $SU(2)\times SU(2)$ . The motivation for the current work is to better understand the physics of the linearized spectrum, both supersymmetric and non-supersymmetric, around the warped deformed conifold and the baryonic branch. Since the PT ansatz can be embedded within our theory, it is clear that the resolved, deformed and singular conifolds can all be found as solutions and there is also some possibility that new solutions may exist within our extended ansatz.

While the five-dimensional, $\N=4$ theory we have constructed is uniquely specified by the number of vector multiplets and the embedding tensor, we also have explicit uplift formulas to ten-dimensional IIB supergravity. This is something which is often quite difficult to obtain, for example a closely related system is the $SU(2)\times U(1)$ gauged supergravity constructed in \cite{Corrado:2002wx} which contains in its solution space the Klebanov-Witten flow \cite{Klebanov:1998hh}, but explicit uplift formula are not available and one is forced to work directly in ten dimensions \cite{Halmagyi:2004jy}. In fact it would be interesting to see if the Heisenberg algebra found here arises as a contraction of the $SU(2)$ gauging in \cite{Corrado:2002wx}. 

There are several direct generalizations of the current work which could provide new insight into the physics of lower dimensional gauged supergravities and string theory. There exists a family of Einstein manifolds related to $T^{1,1}$ called $T^{p,q}$, all of which can be viewed as $U(1)$ fibrations over $S^2\times S^2$ however these do not admit a covariantly constant Killing spinor and thus appear to not preserve any supersymmetry. It would be interesting to determine whether reduction on these manifolds results in a non-supersymmetric theory or a supersymmetric theory with no (canonical) supersymmetric vacuum. Another interesting direction is to consider reductions of IIA on $T^{1,1}$ and compare the resulting embedding tensor of the gauged supergravity to the one found here. By $T$-duality, the spectrum of the ungauged theory is identical to that here, the only difference must lie in the embedding tensor. This will presumably shed some light on the work \cite{Halmagyi:2010st} where evidence was presented that KS cannot have a mirror which is even locally geometric.

Finding a superpotential for this $\N=4$ supersymmetric theory we have presented is a small step further than we have completed in the current work but would be extremely useful and should have direct application to the physics of flux backgrounds. We will return to these issues in a forthcoming publication.

\vskip 1cm
\noindent {\bf \Large Acknowledgements} \\
We would like to thank Davide Cassani, Anatoly Dymarsky, Jerome Gauntlett, Jim Liu, Boris Pioline, Dan Waldram and Nicholas Warner for useful conversations. This work was supported in part by the DSM CEA-Saclay, by the ANR grant 08-JCJC-0001-0, and by the ERC Starting Independent Researcher Grant 240210 - String-QCD-BH.

\begin{appendix}

\section{G-Structures}\label{app:Gstructures}
Here we collate the branching rules required to compute the spectrum of singlets under the $U(1)$ structure group. This provides an alternative way to understand the spectrum in our IIB supergravity ansatz:
\begin{table}[!h]
\begin{center}
{\small 
\begin{tabular}{|c|c|c|}
\hline
field & $SO(5) \ra SO(4)\ra SU(2)_D$& $U(1)$\\
&$ \ra U(1)$ &  neutral field \\
\hline 
$g_{mn}$& ${\bf 15} \ra  {\bf 1}_0+ {\bf 4}+ {\bf 1}_0+ {\bf 9} \ra   {\bf 1}_0+2 \times {\bf 2} + {\bf 1}_0+3\times \bf{3}$& $v, \vbar, u_1, u_2, u_3$\\
&$\ra   {\bf 1}_0+2( {\bf 1}_1+ {\bf 1}_{-1}) + {\bf 1}_0+3( {\bf 1}_0+ {\bf 1}_{1}+{\bf 1}_{-1})$ &\\
\hline
$B_{mn}$&$ {\bf 10} \ra {\bf 4}+{\bf 6} \ra ({\bf 2}+{\bf 2}) + ({\bf 1}_0+{\bf 1}_0+{\bf 1}_0 +{\bf 3}) $& $b, \tb, M_0, \Mbar_0$\\
&$\ra 2({\bf 1}_{1}+{\bf 1}_{-1})  + ({\bf 1}_0+{\bf 1}_0+ {\bf 1}_0+({\bf 1}_0+{\bf 1}_{1}+{\bf 1}_{-1}))$ & \\
\hline
$C_{mn}$&$ {\bf 10} \ra {\bf 4}+{\bf 6} \ra  ({\bf 2}+{\bf 2}) + ({\bf 1}_0+{\bf 1}_0+{\bf 1}_0+{\bf 3}) $& $c, \tc, N_0, \bar N_0$ \\
&$\ra 2({\bf 1}_{1}+{\bf 1}_{-1}) + ({\bf 1}_0+{\bf 1}_0+{\bf 1}_0+({\bf 1}_0+{\bf 1}_1+{\bf 1}_{-1}))$ & \\
\hline
$C_{mnpq}$& ${\bf 5} \ra {\bf 1}_0+{\bf 4} \ra {\bf 1}_0+({\bf 2}+{\bf 2})  $& $k$\\
&$\ra {\bf 1}_0+2({\bf 1}_{1}+{\bf 1}_{-1}) $& \\
\hline
$\phi$ &$ {\bf 1}_0\ra {\bf 1}_0\ra {\bf 1}_0\ra {\bf 1}_0$& $\phi$ \\
\hline
$a$ &$ {\bf 1}_0\ra {\bf 1}_0\ra {\bf 1}_0\ra {\bf 1}_0$& $a$ \\
\hline
\end{tabular}
\caption{\small{Decomposition of scalar fields under the structure group. \label{table:Gdecompsc}}}}
\end{center}
\end{table}

\begin{table}[!h]
\begin{center}
{\small 
\begin{tabular}{|c|c|c|}
\hline
field & $SO(5) \ra SO(4)\ra SU(2)_D$& $U(1)$\\
&$ \ra U(1)$ & neutral field \\
\hline 
$g_{\mu m}$& ${\bf 5} \ra  {\bf 1}_0+{\bf 4} \ra  {\bf 1}_0+({\bf 2}+{\bf 2}) \ra  {\bf 1}_0+2({\bf 1}_1+{\bf 1}_{-1}) $ & $A_1$\\
\hline
$B_{\mu m}$&$ {\bf 5} \ra  {\bf 1}_0+{\bf 4} \ra  {\bf 1}_0+({\bf 2}+{\bf 2}) \ra  {\bf 1}_0+2({\bf 1}_1+{\bf 1}_{-1}) $ & $B_1 $\\
\hline
$C_{\mu m}$& ${\bf 5} \ra {\bf 1}_0+{\bf 4} \ra {\bf 1}_0+({\bf 2}+{\bf 2}) \ra {\bf 1}_0+2({\bf 1}_1+{\bf 1}_{-1})$& $C_1$\\
\hline
$C_{\mu mnp}$& ${\bf 10} \ra {\bf 4}+{\bf 6} \ra  ({\bf 2}+{\bf 2}) + ({\bf 1}_0+{\bf 1}_0+{\bf 1}_0+3)$ & $k_{11}, k_{12}, D_1, \Dbar_1$\\
&$ \ra  2({\bf 1}_1+{\bf 1}_{-1}) + ({\bf 1}_0+{\bf 1}_0+{\bf 1}_0+({\bf 1}_0+{\bf 1}_{1}+{\bf 1}_{-1}))$ & \\
\hline
$B_{\mu\nu}$ &$ {\bf 1}_0\ra {\bf 1}_0\ra {\bf 1}_0\ra {\bf 1}_0$& $B_2$ \\
\hline
$C_{\mu\nu}$ & ${\bf 1}_0\ra {\bf 1}_0\ra {\bf 1}_0\ra {\bf 1}_0$& $C_2$ \\
\hline
\end{tabular}
\caption{\small{Decomposition of form fields under the structure group. \label{table:Gdecompvec}}}}
\end{center}
\end{table}

\section{The One-Forms} \label{app:conventions}
In our ansatz in section \ref{sec:metric} we have used the following standard invariant one-forms on $T^{1,1}$ (see \cite{Minasian:1999tt, Herzog:2001xk}):
\bea
\sig_1&=& c_{\psi/2} d\tha_1 +s_{\psi/2} s_{\tha_1} d\phi_1 \non \\
\sig_2&=& s_{\psi/2} d\tha_1 -c_{\psi/2} s_{\tha_1} d\phi_1 \non \\
\sig_3&=& \half d\psi + c_{\tha_1} d\phi_1 \\
\Sig_1&=& c_{\psi/2} d\tha_2 +s_{\psi/2} s_{\tha_2} d\phi_2 \non \\
\Sig_2&=& s_{\psi/2} d\tha_2 -c_{\psi/2} s_{\tha_2} d\phi_2 \non \\
\Sig_3&=& \half d\psi + c_{\tha_2} d\phi_2 \non,
\eea
where  $c_{\psi/2}=\cos\frac{\psi}{2}$ etc.
\section{Hodge Dualizing} \label{app:hodge}

In signature $(d-1,1)$ the Hodge star squares to
\be
*_d*_d\om_r =(-1)^{1+r(d-r)} \om_r
\ee
where $\om_r$ is an $r$-form. So in $d=5$
\be
*_5*_5 \om_r = -\om_r
\ee
and in $d=10$
\be
*_{10}*_{10}\om_r = (-1)^{1+r} \om_r.
\ee

Here we set up some conventions for embedding the five-dimensional Hodge star in ten dimensions:
\bea
*_{10} \om_p &=&(-1)^p e^{(2-2p)(u_3-u_1)} (*\om_p) \w J'_1 \w J'_2 \w (g_5+A_1) \non \\
*_{10} \Blp \om_p\w (g_5+A_1) \Brp&=& e^{2pu_1 +(8-2p)u_3}(*\om_p) \w J'_1 \w J'_2  \non \\
*_{10} \Blp \om_p\w J'_1 \Brp &=&(-1)^p  e^{(-6+2p)u_1-4u_2 +(2-2p)u_3}(*\om_p) \w  J'_2 \w (g_5+A_1) \non \\
*_{10}\Blp \om_p\w J'_2 \Brp &=&  (-1)^p e^{(-6+2p)u_1+4u_2 +(2-2p)u_3}(*\om_p) \w J'_1  \w (g_5+A_1) \non \\
*_{10} \Blp \om_p\w J'_1 \w (g_5+A_1) \Brp &=& e^{(-4+2p)u_1-4u_2 +(8-2p)u_3}(*\om_p) \w  J'_2 \\
*_{10}\Blp \om_p\w J'_2 \w (g_5+A_1) \Brp &=&e^{(-4+2p)u_1+4u_2 +(8-2p)u_3}(*\om_p) \w J'_1 \non \\
*_{10} \Blp \om_p\w \Om' \Brp  &=& (-1)^p e^{(-6+2p)u_1+(2-2p)u_3}(*\om_p) \w \Om' \w (g_5+A_1) \non \\
*_{10} \Blp \om_p\w \Om' \w (g_5+A_1) \Brp  &=&e^{(-4+2p)u_1+(8-2p)u_3} (*\om_p) \w \Om'\non\\
*_{10} \Blp \om_p\w J'_1 \w J'_2 \Brp &=& (-1)^p e^{(-10+2p)u_1 +(2-2p)u_3}(*\om_p) \w (g_5+A_1) \non 
\eea
where $*_{10}$ is the Hodge star in $d=10$ and $*$ is the Hodge star in $d=5$.

\section{IIB Conventions} \label{conventions}

The $d=10$, IIB action in Einstein frame is\footnote{our conventions are taken from \cite{Polchinski:2000uf} page 12.} \cite{Schwarz:1983qr}
\bea
S_{\IIB}^E&=&\frac{1}{2} \int d^{10}x \Blp \sqrt{G} R-\half  d\phi\w * d\phi -\half e^{2\phi}dC_{(0)}\w * dC_{(0)}
-\frac{e^{-\phi}}{2} H_{(3)}\w * H_{(3)} \non \\
&&-\frac{e^{\phi}}{2} F_{(3)}\w * F_{(3)}-\frac{1}{4} F_{(5)}\w * F_{(5)} -\half C_{(4)}\w H_{(3)} \w F_{(3)} \Brp .
\eea
The full set of equations of motion are
\bea
d(*e^{2\phi}F_{(1)}) &=& -e^{\phi} H_{(3)}\w *F_{(3)} \label{eom1} \\
d (*e^\phi F_{(3)})&=& F_{(5)}\w H_{(3)} \label{eom2} \\
d (*e^{-\phi} H_{(3)})&=&e^{\phi} F_{(1)}\w *F_{(3)} + F_{(3)}\w F_{(5)} \label{eom3} \\
d*d\phi &=&  e^{2\phi} F_{(1)} \w * F_{(1)} - \half e^{-\phi} H_{(3)}\w * H_{(3)} +\half F_{(3)} \w * F_{(3)} \label{eom4} \\
F_{(5)}&=&*F_{(5)} \label{eom5} \\
dF_{(5)} &=& H_{(3)}\w F_{(3)} \label{eom6} \\
R_{MN}&=& \half  \del_M \phi \del_N \phi + \half e^{2\phi} \del_M C_{(0)} \del_N C_{(0)} + \frac{1}{96} F_{MPQRS} F_N^{\ PQRS} \non \\
&& + \frac{e^{-\phi}}{4} \blp H_{M}^{\ PQ}H_{NPQ} -\frac{1}{12}g_{MN} H^{PQR}H_{PQR} \brp   \non \\
&& + \frac{e^{\phi}}{4} \blp F_{M}^{\ PQ}F_{NPQ} -\frac{1}{12}g_{MN} F^{PQR}F_{PQR} \brp  \label{eom7}
\eea

\section{The Five-Form} \label{app:fiveform}
Due to the self-duality in ten dimensions the most involved part of reconstructing the five-dimensional action is the equations of motion for components of the five-form. Here we summarize various steps we have taken. To facilitate computing the exterior derivative, we first write the ansatz \eq{fiveansatz} in terms of untwisted fundamental forms
\bea
F_5&=&e^Z e^{8(u_3-u_1)} \vol_5 + e^Z J_1\w J_2\w (g_5+A_1)  \non \\
&&+K_1\w J_1 \w J_2  -e^{-8 u_1} (*\tK_1) \w (g_5+A_1) \non \\
&& + K_{21}\w J_1 \w (g_5+A_1) + e^{-4u_2+4u_3}(*\tK_{21})\w J_2 \non \\
&& + K_{22}\w J_2 \w (g_5+A_1) +e^{4u_2+4u_3} (*\tK_{22})\w J_1 \non \\
\myvspace
&& + \blp L_{2} \w \Om+c.c \brp \w (g_5+A_1 ) +  e^{4u_3} \blp (*\tL_2)\w\Om +c.c \brp 
\eea
where the unprimed forms are given in terms of the primed ones by
\bea
K_1&=& K'_1 \\
K_{21}&=& K'_{21}-|v|^2 K'_{22} +4 \, \Im\! v L'_2 \\
K_{22}&=&K'_{22} \\
L_2&=&L'_2 - \frac{i\vbar}{2} K'_{22}.
\eea
The Hodge star creats a bit of a mess as well so we have defined some new fields
\bea
\tK_1 &=& K'_1 \non  \\
&=& K_1 \\
\tK_{21} &=& K'_{21} \non \\
&=& K_{21} -|v|^2 K_{22} - 4 \Im \! (vL_2) \\
\tK_{22} &=&  K'_{22} -|v|^2e^{-8u_2} K'_{21} +4e^{-4u_2} \,\Im\! (vL'_2) \non \\
&=& \blp 1+|v|^2 e^{-4u_2} \brp^2 K_{22} -|v|^2 e^{-8u_2} K_{21}  + 4e^{-4u_2}\blp 1+ |v|^2e^{-4u_2}  \brp \Im\! (vL_2)\\
\tL_2 &=&  L'_2  - \frac{i\vbar}{2} e^{-4u_2 }K'_{21}   \non \\
&=& L_2 \blp 1+|v|^2 e^{-4u_2}\brp + \frac{i\vbar}{2} \blp 1+ |v|^2 e^{-4u_2}   \brp K_{22} -\frac{i\vbar}{2}e^{-4u_2} K_{21} -\vbar^2 e^{-4u_2} \Lbar_2
\eea
and then the five-dimensional equations of motion are
\bea
DL_2-3i e^{4u_3} *\tL_2  &=& -M_0 G_3 +N_0  H_3 -  H_2 \w N_1 + M_1\w G_2 \non \\
K_{21}\w F_2+d\blp e^{4(u_2+u_3)}*\tK_{22}  \brp -2 e^{-8u_1}*\tK_1&=&H_{11}\w G_3 +H_3\w G_{11} \label{F5eom6} \\
 K_{22}\w F_2+d\blp e^{4(-u_2+u_3)}*\tK_{21}  \brp -2 e^{-8u_1}*\tK_1&=& H_{12}\w G_3 + H_3\w G_{12} \label{F5eom7} \\
L_2\w F_2 + D(e^{4u_3} *\tL_2) &=& M_1 \w G_3 +    H_3\w N_1 \label{F5eom8} \\
-d\blp e^{-8u_1}*\tK_1\brp &=& H_3\w G_2-H_2\w  G_3 \label{F5eom9}
 \eea
where the two-form $L_2$ is charged
\be
DL_2=dL_2-3iA_1\w L_2.
\ee
The five-dimensional kinetic terms which we get from these equations are rather off-diagonal
\bea
\cL_{F_{(5)},kin}&=& -4 e^{4u_3}\blp 1+|v|^2e^{-4u_2}\brp L_2\w * \Lbar_2   + 4 e^{-4u_2+4u_3}\Blp v^2 L_2\w * L_2 + c.c \Brp \non \\
&& - \half e^{4u_2+4u_3}  \blp 1+|v|^2 e^{-4u_2} \brp^2K_{22}\w * K_{22} - \half e^{-4u_2+4u_3} K_{21}\w * K_{21} \non \\
&& + |v|^2 e^{-4u_2+4u_3}  K_{22}\w *K_{21}  + 2e^{4u_3} \blp 1+ |v|^2 e^{-4u_2}   \brp  \Blp i v K_{22}\w *L_2 +c.c \Brp\non \\
&& -2e^{-4u_2+4u_3}  \Blp iv K_{21}\w *L_2 +c.c \Brp.
\eea
The Bianchi identities are
\bea
d e^Z &=& P(H_{12}-H_{11})+4 \blp M_1 \Nbar_0 - M_0 \Nbar_1 +c.c \brp \label{F5eom1} \\
e^Z F_2 +dK_1+2K_{21}+2K_{22} &=& H_{11}\w G_{12}+ H_{12}\w G_{11}  +  4\blp M_1\Nbar_1 + c.c.\brp \label{F5eom2}\\ 
dK_{21}&=& P\, H_3+H_{11}\w G_2 - H_2\w G_{11} \label{F5eom3}\\
dK_{22}&=&-P\, H_3 + H_{12}\w G_2 -  H_2\w G_{12} \label{F5eom4}
\eea
which we have solved in \eq{F5Bianchi1}-\eq{F5Bianchi4}.

\section{The Three-Forms	} \label{app:threeforms}
The three-forms are given in terms of twisted fundamental forms by
\bea
H_{(3)}&=& H_3 + H_2 \w (g_5+A_1) + H'_{11} \w J'_1 + H_{12}\w J'_2  \non \\
&& +  \Blp M'_1\w \Om' +M_0 \Omp \w (g_5+A_1) +c.c \Brp  \non \\
&& - 4 \, \Im\!(M_0 v)J'_1 \w (g_5+A_1) \label{H3p} \\
F_{(3)}&=& P (J'_1-J'_2)\w (g_5+A_1) +G_3 + G_2 \w (g_5+A_1) + G'_{11} \w J'_1 + G_{12}\w J'_2  \non \\
&& +  \Blp N'_1\w \Om' +N'_0 \Omp \w (g_5+A_1) +c.c \Brp  \non \\
&&-\blp P |v|^2 +4 \, \Im\!(N'_0 v)\brp J'_1 \w (g_5+A_1),
\eea
where
\bea
H'_{11}&=&H_{11}-|v|^2 H_{12}- 4\, \Im \!(v M_1) \non \\
M'_1&=&M_1+\frac{i}{2} \vbar H_{12} \non \\
G'_{11}&=&G_{11}-|v|^2 G_{12}- 4\, \Im \!(v N_1) \\
N'_1&=&N_1+\frac{i}{2} \vbar G_{12} \non \\
N'_0&=&N_0-\frac{i}{2} P \vbar \non.
\eea
The equations of motion are

\bea
 d\blp e^{4(u_1-u_3)-\phi}* H_3\brp&=& -e^Z G_3 + G_2 \w K_1 -G_{12}\w K'_{21}- G'_{11}\w K'_{22} \non \\
&& -4 \blp N'_1\w \Lbar'_2 +c.c \brp +4e^{4u_3} \blp N'_0*\Lbar'_2 +c.c. \brp \non \\
&&+P\blp 1-|v|^2-4\Im(N'_0v)\brp e^{-4(u_2-u_3)} *K'_{21} \non \\
&&- Pe^{4u_2+4u_3)} *K'_{22} + e^{4(u_1-u_3)+\phi} da\w (*_5 G_3) \\
&& \non \\
d\blp e^{4(u_1+u_3)-\phi} * H_2 \brp &=& 2 (1-|v|^2)e^{-4(u_1+u_2)-\phi} *H'_{11} +2 e^{-4(u_1-u_2)-\phi} *H_{12}  \non \\
&& +e^{4(u_1-u_3)-\phi} *H_3\w F_2 +8 e^{-4u_1-\phi} \Im\!(v *M'_1) \non \\
&& + G_3\w K_1 + e^{-4(u_2-u_3)}G'_{11} \w *K'_{21} + e^{4(u_2+u_3)}G_{12} \w *K'_{22} \non \\
&&+4e^{4u_3}\blp N'_1\w* \Lbar'_2 +c.c.\brp + e^{4(u_1+u_3)+\phi}da\w * G_{2} \label{H2eom}\\
&& \non 
\eea
 \bea
 d\blp  e^{-4(u_1-u_2)-\phi} * H_{12}  \brp&=&  - 4e^{-4u_1-\phi}\Im\!(*M'_1\w Dv)+12 e^{-4u_1+8u_3-\phi}\Re\!(M_0 v) \vol_5\non \\
 && +\blp P(1-|v|^2)-4\Im(N'_0 v)\brp e^Z e^{8(u_3-u_1)} \vol_5 + e^{-8u_1} G'_{11}\w *K_1 \non \\
&&+ e^{4(u_2+u_3)} G_{2}\w *K'_{22}  -G_3 \w K'_{21} + e^{-4(u_1-u_2)+\phi}da\w * G_{12}\label{H12eom} \\
&& \non \\
\non \\
d\blp e^{-4(u_1+u_2)-\phi} * H'_{11} \brp &=& -P e^Z e^{8(u_3-u_1)} \vol_5+e^{-8u_1} G_{12}\w *K_1 + e^{-4(u_2-u_3)} G_{2}\w *K'_{21}  \non \\
&& -G_3 \w K'_{22} + e^{-4(u_1+u_2)+\phi}da\w * G'_{11}\label{H11eom}  \\
&&\non \\
D\blp e^{-4u_1 -\phi}* M'_1 \brp &=& -3i e^{-4u_1+8u_3-\phi} M_0\vol_5   \non \\
&&+6 e^{-4u_1-4u_2+8u_3-\phi} \vbar \Im \!( vM_0) \vol_5 -\frac{1}{2i}e^{-4(u_1+u_2)-\phi} *H'_{11} \w  D\vbar \non \\
&& e^Z e^{8(u_3-u_1)}N'_0\, \vol_5 + e^{-8u_1}N'_1\w *K_1 \non  \\
&&- G_3\w L'_2 +e^{4u_3} G_2\w *L'_2 + e^{-4u_1+\phi} da\w  *N'_1\label{M1eom} \\
&&\non \\
d\blp e^{4(u_1-u_3)+\phi}* G_3 \brp &=&  e^Z H_3 -   K_1\w H_2 +  K'_{22}\w H'_{11} + K'_{21}\w  H_{12}   \non \\
&&  +4 e^{-4u_2+4u_3} \Im\!(vM_0)*K'_{21}+4 \blp L'_2\w \Mbar'_1 +c.c. \brp\non \\
&&-4e^{4u_3} \blp *L'_2 \Mbar_0+c.c.\brp  \\
&&\non\\
d\blp e^{4(u_1+u_3)+\phi} * G_2 \brp&=& e^{4(u_1-u_3)+\phi} *G_3\w F_2    +2 (1-|v|^2) \blp e^{-4(u_1+u_2)+\phi} *G'_{11} \brp \non \\
&&+2e^{-4(u_1-u_2)+\phi} *G_{12}-8 e^{-4u_1} \Im \!(v*N'_1)+ K_1\w H_3\non \\
&&-4e^{4u_3}\blp M'_1\w *\Lbar'_2 +c.c.\brp +e^{-4u_2+4u_3} *K'_{21}\w H'_{11} \non \\
&&+e^{4u_2+4u_3} *K'_{22}\w H_{12} \\
&&\non\\
d\blp  e^{-4(u_1-u_2)+\phi} * G_{12}  \brp&=& -4e^{-4u_1+\phi} \Im\!(*N'_1\w Dv)+12 e^{-4u_1+8u_3+\phi} \Re\!(N'_0v) \vol_5 \non \\
&&K'_{21} \w H_3-  e^{4(u_2+u_3)} *K'_{22}\w H'_2 - e^{-8u_1} *K_1\w H'_{11} \non \\
&&+4 \Im\!(M_0 v) e^Z e^{8(u_3-u_1)}  \\
&&\non\\
d\blp e^{-4(u_1+u_2)+\phi} * G'_{11} \brp&=&  K'_{22}\w H_3  -e^{-4u_2+4u_3} *K'_{21}\w H_2- e^{-8u_1}*K_1\w H_{12}  \\
&&\non \\
\non \\
 D\blp e^{-4u_1 +\phi}* N'_1 \brp &=&- 3i e^{-4u_1+8u_3+\phi} N'_0\vol_5  -\frac{1}{2i}e^{-4u_1-4u_2} (*G'_{11})\w D\vbar \non \\
&& +\frac{3\vbar}{2} e^{-4u_1 -4u_2+8u_3+\phi} \blp P |v|^2+4 \Im\!(N'_0 v) \brp\vol_5  \non \\
&& +L'_2 \w H_3 - e^{4u_3}*L'_2\w H_2   - e^{-8u_1} *K_1\w M'_1\non \\
&&- M_0e^Z e^{-8(u_1-u_3)} \vol_5
\eea

\section{Dilaton-Axion}

For the dilaton axion equations of motion is necessary to construct the follows quantities:
\bea
F_{(1)}\w *_{10}F_{(1)}&=&J'_1\w J'_2 \w (g_5+A_1) \Bslb da \w * da \Bsrb \\
 H_{(3)}\w *_{10}H_{(3)}&=&J'_1\w J'_2\w g_5\w \Bslb e^{4(u_1-u_3)} H_3\w * H_3 +e^{4(u_1+u_3)} H_2\w * H_2 \non \\
&&+ e^{-4(u_1+u_2)} H'_{11}\w * H'_{11} +  e^{-4(u_1-u_2)} H_{12}\w * H_{12}  \non \\
&&+ 8 e^{-4u_1 } M'_1\w * \Mbar'_1 +8 e^{-4u_1+8u_3}|M_0|^2 \vol_5   \non \\
&& +16e^{-4u_1-4u_2+8u_3}\slb\Im\!(M_0 v)\srb  ^2 \vol_5  \Bsrb   \\
F_{(3)}\w *_{10} F_{(3)}&=& J'_1\w J'_2\w (g_5+A_1)\w \Bslb e^{4(u_1-u_3)}G_3\w * G_3 +e^{4(u_1+u_3)}G_2\w * G_2 \non \\
&& + e^{-4u_1-4u_2} G'_{11}\w * G'_{11} + e^{-4u_1+4u_2} G_{12}\w * G_{12} + 8 e^{-4u_1 } N'_1\w * \Nbar'_1 \non\\
&& +e^{-4u_1+8u_3} \Blp 8 |N'_0|^2  +e^{4u_2} P^2+e^{-4u_2}  \blp P(|v|^2-1)+4\, \Im\!(N'_0 v) \brp^2\Brp \vol_5 \Bsrb\non \\
 H_{(3)}\w *_{10}F_{(3)} &=& J'_1\w J'_2\w (g_5+A_1)\w  \Bslb e^{4(u_1-u_3)} H_3\w *G_3 +e^{4(u_1+u_3)} H_2\w *G_2 \non \\
&& +e^{-4(u_1+u_2)} H'_{11}\w *G'_{11}  +e^{-4(u_1-u_2)} H_{12}\w *G_{12}  \non \\
&&+4e^{-4u_1} \blp M'_1\w * \Nbar'_1 +c.c. \brp +\Blp 4 e^{-4u_1+8u_3}\blp M_0\Nbar'_0+c.c. \brp \non \\
&&-4e^{-4u_1-4u_2+8u_3}\,\Im\!\!(M_0v)\blp P(1- |v|^2)-4\Im\!\!(N'_0v)\brp\Brp\vol_5 \Bsrb
\eea
then \eq{eom1} and \eq{eom4} give the relevant equations of motion.

\end{appendix}


\providecommand{\href}[2]{#2}\begingroup\raggedright\endgroup

\end{document}